\documentclass[aps,pre,preprint,groupedaddress]{revtex4}
\usepackage{graphicx,amssymb}
\usepackage{amsmath}
\usepackage{psfig}
\usepackage{epsfig}
\makeatother      
\newlength{\newwidth}
 
\begin{document}

\preprint{}

\title{Shock formation in an
exclusion  process with creation and annihilation}


\author{M. R. Evans$^1$, R. Juh{\'a}sz$^2$, L. Santen$^2$}
\affiliation{$^1$School of Physics, University of Edinburgh,
Mayfield Road, Edinburgh, EH9 3JZ, United Kingdom\\}

\affiliation{$^2$Theoretische Physik, Universit\"at des Saarlandes,
66041 Saarbr\"ucken, Germany}


\date{\today}

\begin{abstract}
We investigate  shock formation in an asymmetric exclusion  process with
creation and annihilation of particles in the bulk.
We show 
how the continuum mean-field equations can be studied
analytically and hence derive   the phase diagrams of the model.
In the large system-size limit
direct simulations of the model show that the stationary
state  is correctly described by the 
mean-field equations, thus the predicted mean field
phase diagrams are expected to be exact. The emergence of shocks and 
the structure of the phase diagram are discussed.
We also analyse the fluctuations of the shock position by using 
a phenomenological random walk picture of the shock dynamics. 
The stationary distribution of shock positions is calculated, by
virtue of which
the numerically determined finite-size scaling
behaviour of the shock width is explained.
\end{abstract}

			      \maketitle

\newcommand{\del}{\partial}
\newcommand{\ul}{\underline}
\newcommand{\im}{{\rm i}}
\newcommand{\be}{\begin{eqnarray}}
\newcommand{\ee}{\end{eqnarray}}
\newcommand{\ba}{\begin{array}}
\newcommand{\ea}{\end{array}}
\newcommand{\f}[2]{\frac{#1}{#2}}
\newcommand{\mc}{\mathcal}
\newcommand{\bs}{\backslash}
\newcommand{\bra}[1]{\langle{#1}\!\mid}
\newcommand{\ket}[1]{\mid\!{#1}\rangle}
\newcommand{\D}{\ensuremath{\mathrm{d}}}


\section{Introduction}
\label{intro}
The analysis of  self-driven particle-models is a central issue of
non-equilibrium statistical mechanics
\cite{Liggett,schutzreview,privman,chowd,hinrichsen}.  
 These models show a variety 
of generic non-equilibrium effects, in particular in  low dimensions. 
Although the behaviour  is generally
rather complex, some models  
are simple enough to be analysed in great detail. In rare cases it 
is even possible to obtain exact results for the stationary state
\cite{hinrichsen,schutzreview}. 
A very prominent example of such a model is the asymmetric 
exclusion process (ASEP) which comprises
particles hopping in a preferred direction
under the constraint that they cannot occupy
the same lattice site.
Exact solutions for the stationary state 
exist for 
periodic \cite{Liggett,schadschneider} and open boundary conditions
\cite{DEHP,SD}  
and different update schemes~\cite{ERS,dGN,rajewsky}.
Therefore this model has been used in order to develop more general
concepts for systems far from equilibrium, e.g. a free-energy
formalism \cite{dls}. Variants 
of the   model have also
allowed the study  of shocks which are discontinuities in the density of particles over a microscopic distance \cite{JL}.

The ASEP and related  models are not only of academic interest, but
have a number of important applications, e.g. as  simplified traffic 
models~\cite{chowd}. Here we are interested in a variant of the ASEP,
which is
motivated by biological transport processes in living cell systems,
where particle non-conservation in the bulk of the system is allowed
\cite{parma} (see also \cite{LKN}).

An important feature  of living cells is their ability to move and
to generate forces~\cite{alberts_book,howbook,leshouches}. On a
microscopic level these forces  
are mostly generated by motor-proteins~\cite{howard,juelrmp}, which are able 
to perform directed motion along one-dimensional paths or filaments.

Many different motor proteins can  be distinguished~\cite{howard},
but they have a few common properties:
a head, 
which can couple to a filament, where it performs a directed 
stochastic motion; and a tail, which  is attached to a specific 
load, which has to  be transported through the cell. The coupling
of the motor protein heads to the filament is reversible,
thus the motor proteins will attach to the filament,
perform stochastic directed motion for some
time and eventually  detach from the filament.
The typical distance between attachment and detachment of
the motor protein to the filament depends on the 
particular type of filament. 

To a first approximation the motion of many motor proteins along a
filament can be modelled by the asymmetric exclusion process.  An
important feature, which is not described by the ASEP, is the
attachment and detachment of the motor-protein heads. This feature has
been included in a recent model by Parmeggiani {\it et al}
\cite{parma} (see also \cite{LKN}), which can be viewed as a grand-canonical counterpart of
the ASEP in the sense that in the bulk the particle number is not
conserved.

For open boundary conditions, however, the situation is different. In
this case, if the rates of attachment scale correctly with system
length, one observes a subtle interplay between the left, right and
bulk particle reservoirs.  If we fix the attachment and detachment
rates then there are whole regions of the phase diagram (spanned by
the densities of the boundary reservoirs) where one observes the {\em
localisation} of shocks in the bulk of the system. This is in contrast
to the ASEP, where shocks move with  constant velocity and  are
generally driven to the boundary of an open system.  The shock has
zero velocity only on the phase boundary where two phases of different
density co-exist.  Even in this case the shock
is not localised since it diffusively explores
the whole system.

The physical origin of the shock localisation in the present model and
a discussion of the phase diagram, in particular the phase where
shocks appear, are the subjects of this article.  We shall show how
the phase diagram can be predicted through simple considerations
pertaining to a continuum mean field description which only retains
first order terms i.e. the phase diagram can be predicted through the
study of a simple first-order nonlinear partial differential equation.
Going beyond the mean level, in the shock phase we describe the
dynamics of the shock by a random walker with space-dependent hopping
rates. In this way the localisation of the shock can be understood.

The article is organised as follows. In the next section we give the
definition of the model and introduce the stationary solution in case
of periodic boundary conditions.  In section~\ref{mean} we introduce
the mean field equations for the open system and discuss their
solution by means of characteristics. Then we discuss the stationary
solutions on the mean field level and compare the mean field results
to Monte Carlo simulations.  Fluctuations of the shock positions are
analysed in section~\ref{shock} and finally some concluding remarks
are given in the last section.
 
\section{Model} 
\label{model}
We consider a one-dimensional open chain of $N$ sites, which 
can either be empty ($\tau_i = 0$) or be occupied with one particle 
($\tau_i = 1$). Particles 
can jump to the neighbouring site if it is empty. In addition,
the bulk sites are coupled to particle reservoirs, i.e. 
particles are attached with rate $\omega_A$ and deleted with 
rate  $\omega_D$. The particle reservoirs at the boundaries 
 are different from 
the bulk. At the first site particles are attached with rate
$\alpha$ and deleted with rate $\omega_D$, while at the last site 
particles attach with rate $\omega_A$  and are  deleted  
with rate $\beta$~\cite{parma}. By a rate $r$ it is meant that
in infinitesimal time interval
$dt$ the probability of the event occurring is $rdt$.
Schematically the dynamics can be written as follows:
\be 
{*}\, 0 &  \stackrel{\omega_A}{\longrightarrow}& *\, 1\\ 
1\, *  & \stackrel{\omega_D}{\longrightarrow} & 0\, *\\
1\, 0 & \stackrel{1}{\longrightarrow}& 0\, 1
\label{bulk_update}
\ee
where $1(0)$ corresponds to an empty (occupied) site and 
$*$ implies that the update is independent of the state.

The dynamics at the left hand boundary (site 1) is
insertion of particles
\be 
0 & \stackrel{\alpha}{\longrightarrow}& 1 
\label{lh}
\ee
and at the right hand boundary (site $N$) is removal of particles
\be
1 & \stackrel{\beta}{\longrightarrow} & 0 \ .
\label{rh}
\ee

The exact equation for the evolution
of the particle densities $\langle \tau_i \rangle$ 
away from the boundaries ($1<i<N$)  is given by
\begin{equation}
\frac{\D \langle \tau_i \rangle}{\D t}= \langle \tau_{i-1}(1 - \tau_i)
\rangle - \langle \tau_i (1 -\tau_{i+1}) \rangle 
+\omega_A \langle 1-\tau_i \rangle - \omega_D \langle \tau_i \rangle
\;,
\label{Ev1}
\end{equation}
where $\langle \dots \rangle$  denotes the statistical average.
At the boundaries the densities evolve as: 
\begin{eqnarray}
\frac{\D \langle \tau_1 \rangle}{\D t}&= &
 - \langle \tau_1 (1 -\tau_{2}) \rangle 
+\alpha \langle 1-\tau_1 \rangle - \omega_D \langle \tau_1 \rangle
\;,\\
\frac{\D \langle \tau_N \rangle}{\D t}&= &
  \langle \tau_{N-1} (1 -\tau_{N}) \rangle 
+\omega_A \langle 1-\tau_N \rangle - \beta \langle \tau_N \rangle
\;.
\label{eq:bound}
\end{eqnarray}

First consider the steady state
on a periodic system (where
site $N+1$ is identified with site 1 and (\ref{lh},\ref{rh}) do not apply).
Assuming translational invariance, (\ref{Ev1}) is satisfied by
$\langle \tau_i \rangle =\rho_{\rm eq}$ where
\be
\rho_{\rm eq} = \omega_A/(\omega_A + \omega_D)\;.
\label{LD}
\ee
We refer to the density (\ref{LD}) as the equilibrium density
as it is the density obtained in the Langmuir absorption model
\cite{Davis}.
Furthermore, it can be verified that the steady state of the system 
is given by a product measure with density of particles $\rho_{\rm eq}$.

We can apply particle-hole symmetry in order to simplify the
discussion of the model. In the case of open boundary conditions,
the system of equations
(\ref{Ev1}--\ref{eq:bound})  is invariant under the
simultaneous exchanges  $\alpha \leftrightarrow \beta$, $\omega_A
\leftrightarrow \omega_D$, $i \leftrightarrow L-i$ and $ \rho_i
\leftrightarrow 1-\rho_i$.

\section{Mean field equations and Characteristics}
\label{mean}
In the large $N$ limit we can make the continuum mean field approximation
to  (\ref{Ev1}). First we factorise correlation functions
by replacing
$\langle \tau_i (1 -\tau_{i+1}) \rangle$ with $\langle \tau_{i-1}\rangle
(1-\langle \tau_i \rangle)$ \cite{DDM}
then we set 
\be
\langle \tau_{i\pm1} \rangle
 = \rho(x) \pm \frac{1}{N} \frac{ \partial \rho}{\partial x}
+\frac{1}{2N^2} \frac{ \partial^2 \rho}{\partial x^2} \ldots\;.
\label{eq:dev}
\ee
Keeping leading order terms in $1/N$,
one obtains
\be
\frac{ \partial \rho}{\partial \tau}=
-(1-2\rho) \frac{ \partial \rho}{\partial x}
+\omega_D N \left[K -(1+K)\rho \right] \ ,
\label{diffeq}
\ee
where $\tau= t/N$ and 
\begin{equation}
K = \omega_A /\omega_D\;.
\label{Kdef}
\end{equation}
For the open system we shall be interested in the scaling limit
where
\be
\Omega_A = \omega_A N \qquad \textrm{and}  \qquad \Omega_D = \omega_D N 
\ee
are finite as $N \to \infty$. The boundary conditions become
$\rho(x=0)=\alpha$    and $\rho(x=1)=(1-\beta)$.

To understand the first-order  differential equation
(\ref{diffeq}) one can study the characteristics \cite{Debnath},
which are  defined for a quasi-linear equation, 
\be
a(x,\tau,\rho)
\frac{ \partial \rho}{\partial \tau}
+b(x,\tau, \rho) \frac{ \partial \rho}{\partial x}
= c(x,\tau,\rho)\;,
\label{qleq}
\ee
by the equations
\be
\frac{ d x}{d \tau}= \frac{b(x,\tau,\rho)}{a(x,\tau,\rho)}\qquad
\frac{ d\rho}{d \tau}= \frac{c(x,\tau,\rho)}{a(x,\tau,\rho)}\;.
\ee
Roughly speaking, characteristics are curves along which information
about the solution
propagates from the boundary conditions of the partial differential
equation.
Based on characteristics, Lighthill and Whitham \cite{LW} developed
the theory of kinematic waves for equations with mass conservation
and showed how kinematic shock waves arise. Here we generalise this picture to 
equation (\ref{diffeq})
where the number of particles is not conserved.

In the present case the characteristics are given by
\be
\frac{ \partial x}{\partial \tau}&=& 1-2\rho
\label{Char1}\\
\frac{ \partial \rho}{\partial \tau}&=& \Omega_D \left(
K -(1+K)\rho\right) \ .
\label{Char2}
\ee 
Equations
(\ref{Char1},\ref{Char2}) are to be interpreted as
kinematic waves \cite{LW}, which propagate changes in the density, moving with
speed $1-2\rho$ but with the density of the wave $\rho$ itself
changing with time.  In the absence of creation and annihilation of particles
the
density of the wave is constant in time 
and the waves propagate in straight lines.
However in the presence of creation and annihilation of particles
the waves will
follow a curve in the $x$--$t$ plane. 
For example,  consider  a density fluctuation starting at the left boundary
with
$\rho(0) =\alpha <1/2$ and $\alpha < K/(1+K)$. 
Initially the fluctuation will propagate to the right
with speed $1-2\rho$ with its density increasing
and speed decreasing. If $K/(1+K) <1/2$ i.e. $K < 1$
the density will approach
 $\rho = K/(1+K)$ and the fluctuation  will
 propagate to the right with a fixed speed. 
However if $K/(1+K) >1/2$, after some time
the density of the fluctuation will reach $\rho =1/2$
and the fluctuation will cease to propagate.
Similarly, a kinematic wave starting at the right boundary with
$\rho(1)=1-\beta$ where $\beta <1/2$ and $\beta <1/(1+K)$
will travel to the  left with decreasing density and decreasing
speed.

When two characteristic lines cross, multivalued densities are implied,
therefore the description (\ref{diffeq}) by a first order differential
equation breaks down. However Lighthill and Whitham showed that the
effect is that a shock, i.e.  a discontinuity between the densities
$\rho_1$ and $\rho_2$, arises at the meeting points of the two
characteristics and this discontinuity travels with speed $v_s$. This
speed is determined by balance of mass current to be \cite{LW} 
\be v_s
= \frac{\rho_2(1-\rho_2)-\rho_1(1-\rho_1) }{\rho_2 - \rho_1}
=1-\rho_1-\rho_2 \;.
\label{vshock}
\ee

In the present case, although the mass is not conserved, the mass
current between through the shock still implies that its velocity is
given by (\ref{vshock}). Thus for $\alpha<1/2$, $\beta<1/2$ we have
the possibility of a shock forming then being driven to a position
where the mass current through it is zero and the shock remains
stationary.  In the next section we shall show how this picture is
borne out by solving the steady state mean-field equation.

\section{Steady State Solution}
\label{steady}
Setting the time derivative of
(\ref{diffeq}) to zero yields
\be
(1-2\rho) \frac{ \partial \rho}{\partial x}
-\Omega_D\left[K -(1+K)\rho\right]=0
\label{stst}
\ee
This is a first order  ordinary  differential equation,
which in principle can be solved analytically. 
The only difficulty is the occurrence of shocks in the solution. To
construct the solution
we integrate from the left boundary ($\rho(0)=\alpha$) to find a profile
 $\rho_l(x)$:
\be
x &=& \frac{1}{\Omega_D} \int_{\alpha}^{\rho_l(x)}d\rho
\frac{1-2\rho}{K -(1+K)\rho}\\
&=&
\frac{1}{\Omega_D(1+K)}
\left[ 2(\rho_l-\alpha) + \frac{K-1}{(1+K)}
\ln \left| \frac{K-(1+K)\rho_l}{K-(1+K)\alpha}\right|\, \right]\; ;
\label{lint}
\ee
and integrate  from the right boundary ($\rho(1)= 1-\beta$) to find
a profile
$\rho_r(x)$ through
\be
1-x =
\frac{1}{\Omega_D(1+K)}
\left[ 2(1-\beta -\rho_r) + \frac{K-1}{(1+K)}
\ln \left| \frac{K-(1+K)(1-\beta)}{K-(1+K)\rho_r}\right|\, \right]\;.
\label{rint}
\ee
To determine the full profile across the system we have 
generally to match these two profiles at a shock whose position is to be determined.

\subsection{The case  $K=1$}
First we consider
the  special case of $K=1$, i.e. where the attachment and detachment
processes have equal rates.  The steady-state mean-field equation (\ref{stst})
reads 
\be
(2\rho -1)({\partial \rho \over \partial x}-\Omega)=0,
\label{lin}
\ee
where $\Omega \equiv \Omega_A=\Omega_D$.
This  can be solved 
explicitly by a piecewise linear trial function leading to the condition:
either
$\rho(x) = \mbox{const} =\rho_{\rm eq} = {1\over2}$  or 
$\rho (x)$ has slope $\Omega$.
Note that this solution implies that higher
derivatives in eq.~(\ref{eq:dev}) vanish exactly.

\begin{figure}[h]
\centerline{\epsfig{figure=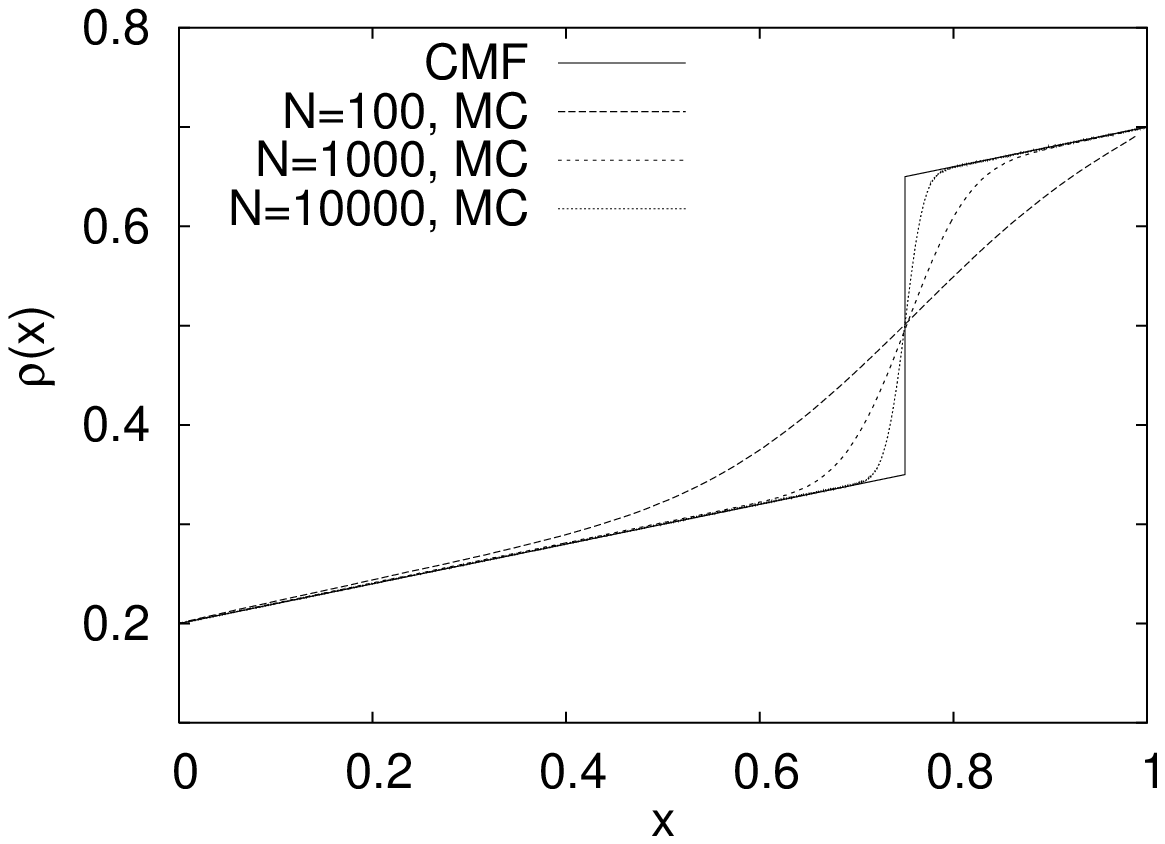,width=0.48\linewidth}\epsfig{figure=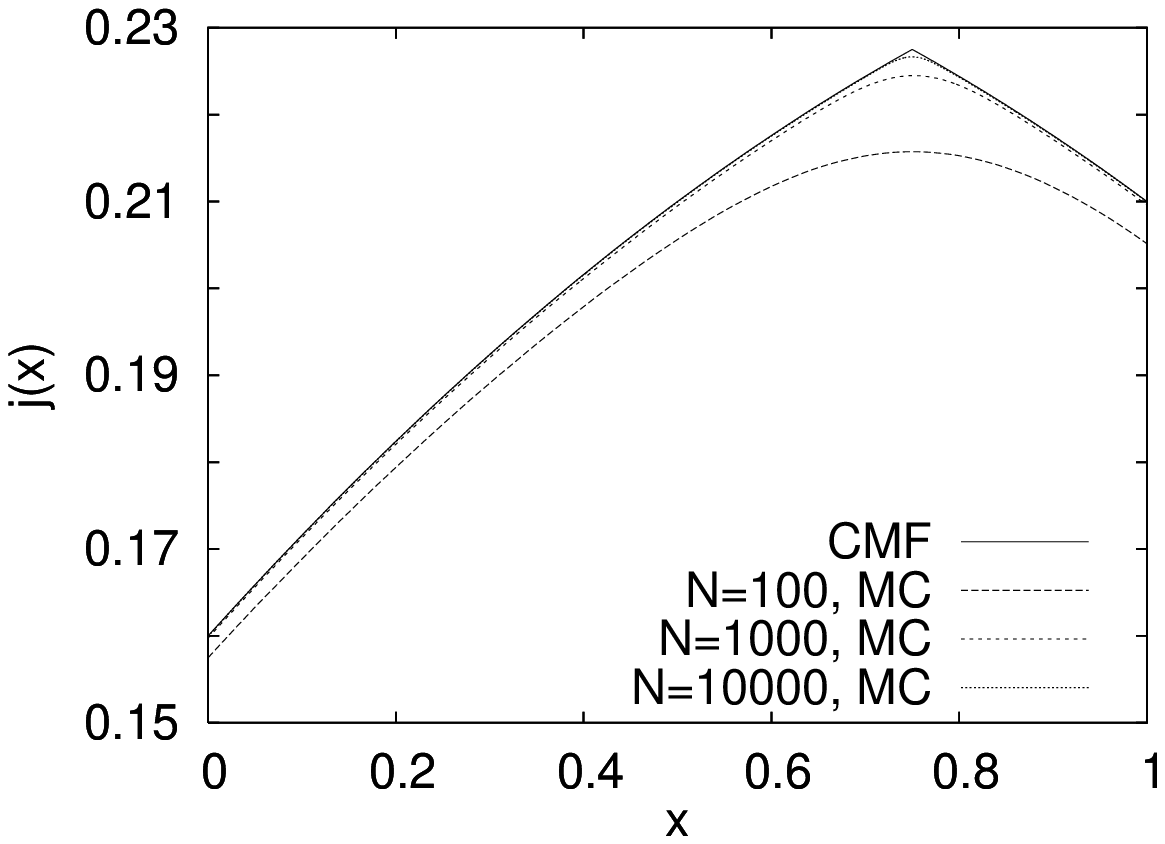,width=0.48\linewidth}}
\caption{ Density  $\langle \tau_i \rangle$ (left) and flow 
$\langle j_i \rangle = \langle  \tau_i (1- \tau_{i+1} ) \rangle$ 
 (right) profiles for $\alpha=0.2$,
  $\beta=0.3$, $K=1$, and $\Omega = 0.1$. For the MC results we set 
  $x = i/N$. The continuum mean-field theory results are compared
  to Monte Carlo (MC) simulations of different system sizes. The 
  agreement between MF and MC results is improved for larger 
  system sizes. }
\label{fig_abI}
\end{figure}

\begin{figure}[h]
\centerline{\epsfig{figure=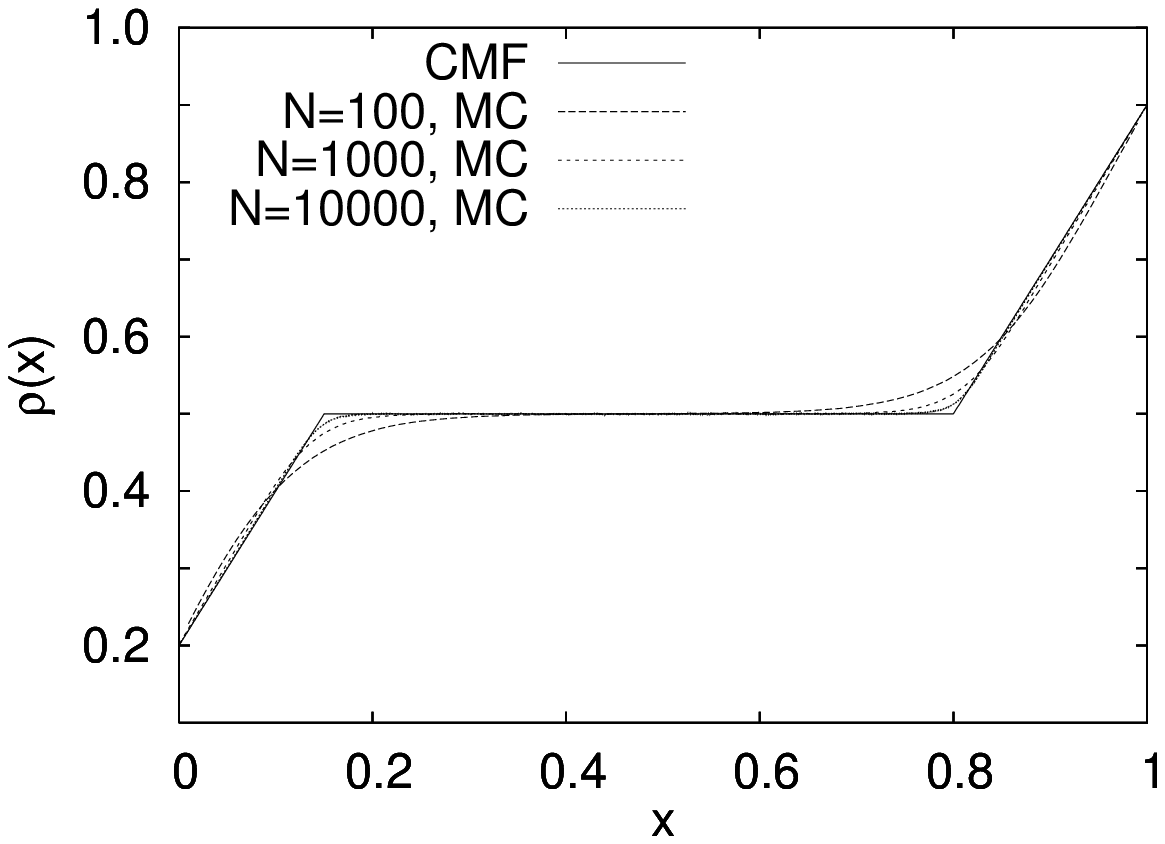,width=0.48\linewidth}\epsfig{figure=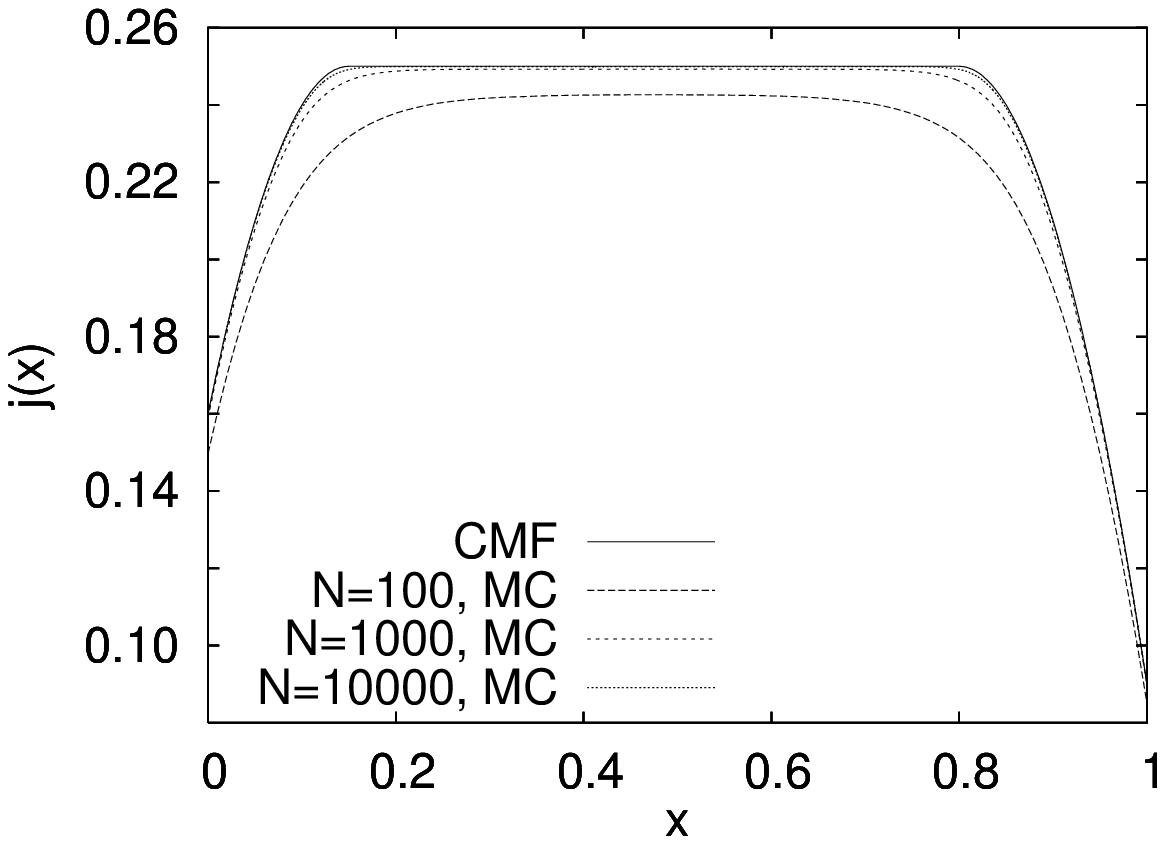,width=0.48\linewidth}}
\caption{Same as fig.~\ref{fig_abI}, but the parameters  $\alpha=0.2$,
  $\beta=0.1$, $K=1$ and $\Omega = 2.0$ are used.  For this larger 
  value of  $\Omega = 2.0$ shock formation is preempted by reaching
  the density   $\rho_{\rm eq} = 1/2$ see (\ref{LD}).  }
\label{fig_abplateau}
\end{figure}

First we consider the parameter regime $\alpha < 1/2$, $\beta  < 1/2$, 
where shock formation is possible.
In order for the shock's position to be
stable i.e. for it not to be driven out of the system
the shock's speed, as determined by (\ref{vshock}), should be zero.
Thus the densities at the discontinuity should be related by
$\rho_r (x_s) = 1-\rho_l(x_s)$ and this determines the position of the shock
$x_s$:
\be
x_s = \frac{ \beta - \alpha  }{2 \Omega} + \frac{1}{2}.
\ee
The height of the shock $\Delta$ is given by 
\be
\Delta = \rho_r(x_s) - \rho_l(x_s) = 1 - (\alpha + \beta) - \Omega
\equiv \Omega_c - \Omega \ . 
\ee
The last equation can be used in order to discuss the parameter
dependence of the model. If $\Delta$ is  positive and $0 <x_s<1$, 
we find indeed 
a shock, which connects two domains with linear density
dependence, as illustrated in fig.~\ref{fig_abI}. For $\Omega >
\Omega_c$ one does not observe shock formation, but a section
 $ \frac{1-2\alpha}{2 \Omega} < x <1- \frac{1-2\beta}{2 \Omega}$, where
$\rho =  \frac{1}{2}$ (see fig.~\ref{fig_abplateau}).

In the ASEP the line  $\alpha =
\beta$ is  a phase boundary where shocks, between a high density region coexisting
with a low density region, exist:
Mean field theory predicts the shock is at $x=1/2$
although the exact solution shows that the shock is actually delocalised
and yields a linear density profile \cite{DEHP}.
In the present case,
although the shock's position is $x=1/2$ when $\alpha =\beta$, the role of this line changes
and one does not observe a phase transition in crossing it.

Moreover linear density profiles, which are observed for 
$\alpha + \beta + \Omega = 1$, do not signify phase
coexistence as for the ASEP, but indicate a vanishing height of the 
shock ($\Delta = 0 $). Also note that a strictly linear profile 
is observed only in the limit $N\to \infty$ in contrast to the 
the disorder line of the ASEP (which is recovered in case of  
$\Omega=0$) where the constant density solution is valid for finite
systems as well.

\begin{figure}[h]
\centerline{\epsfig{figure=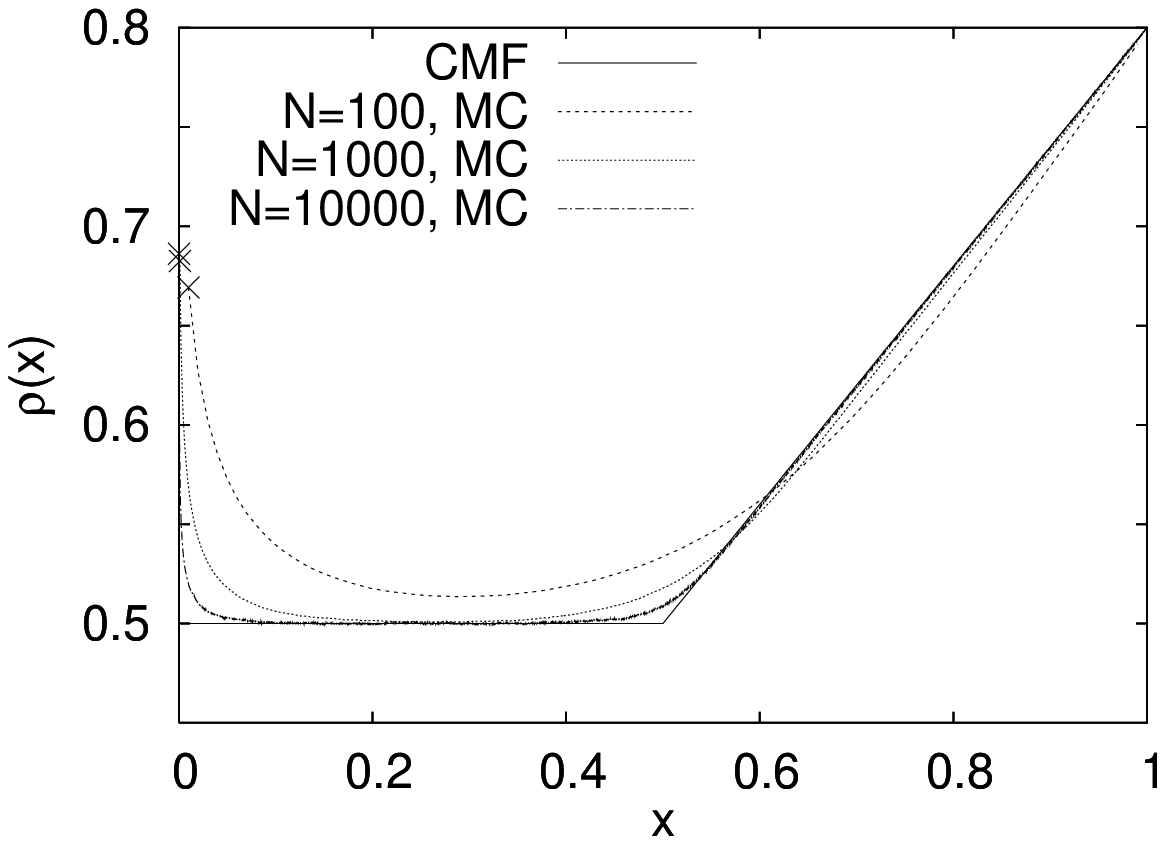,width=0.48\linewidth}\epsfig{figure=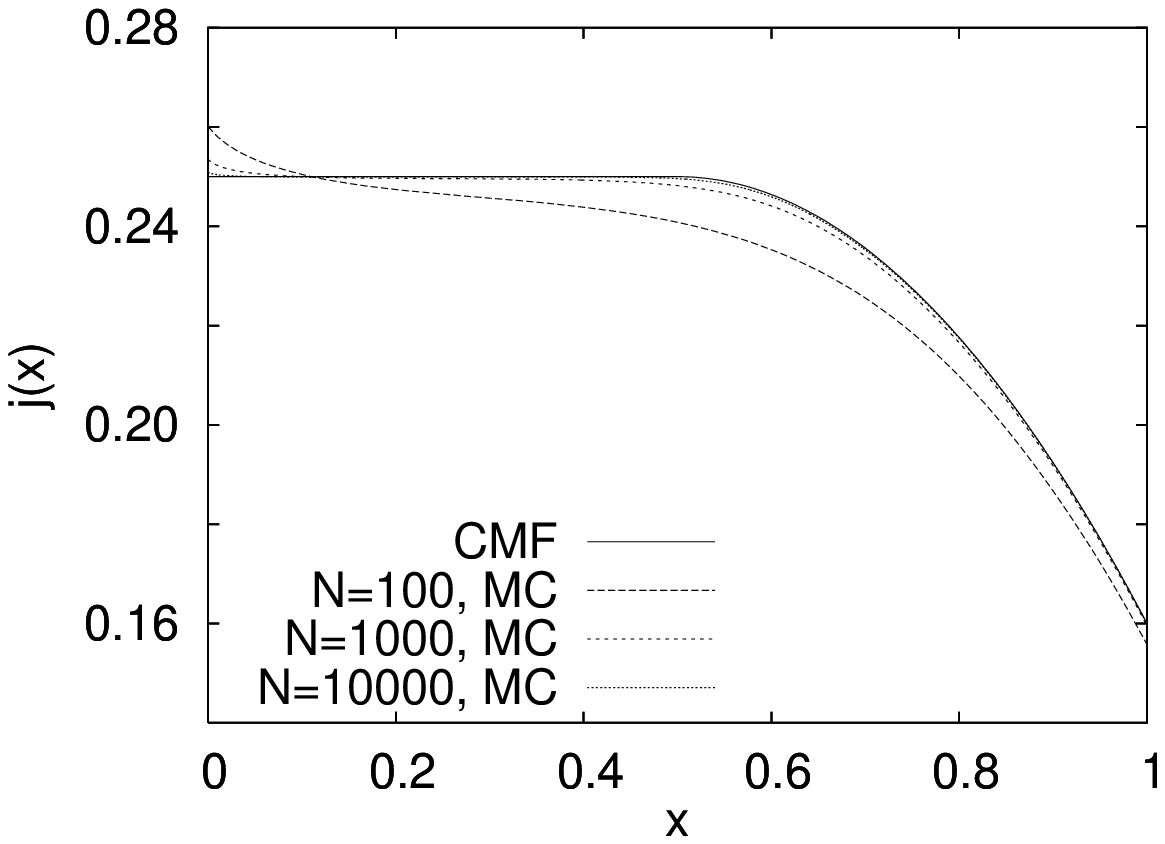,width=0.48\linewidth}}
\caption{Same as fig.~\ref{fig_abI}, but the parameters  $\alpha=0.8$,
  $\beta=0.2$, $K=1$ and $\Omega = 0.6$ are used. The average
  densities at the first site are indicated by symbols.}
\label{fig_abIII}
\end{figure}

Next we discuss the case $\alpha > 1/2$ and $\beta < 1/2$.  In this 
parameter regime the  density profile is given by
\be 
\rho(x) = \max\left[ 1-\beta - \Omega (1-x), \frac{1}{2} \right] 
\qquad \textrm{ for } x>0.
\label{K1H}
\ee  
Equation (\ref{K1H}) implies a singularity at $x=0$
since the boundary condition is $\rho(0)=\alpha$.
For finite systems the singularity at $x=0$
is softened as shown in 
figure fig.~\ref{fig_abIII}. The behaviour for $\alpha > 1/2$ and
$\beta < 1/2$ can be obtained from particle-hole symmetry. 

Finally in the maximal current regime the density profile in the
limit $L\to \infty$ is  given  by
 
\begin{equation}
\rho(x) = 
\begin{cases} 
\alpha & \text{for  $x =0$} \\
1/2 & \text{ for } 0< x < 1 \\
1-\beta & \text{ for } x = 1.
 \end{cases} 
\end{equation}

i.e., apart from the two singular points $x=0$ and $x=1$, the density 
is constant.

\begin{figure}[h]
\centerline{\epsfig{figure=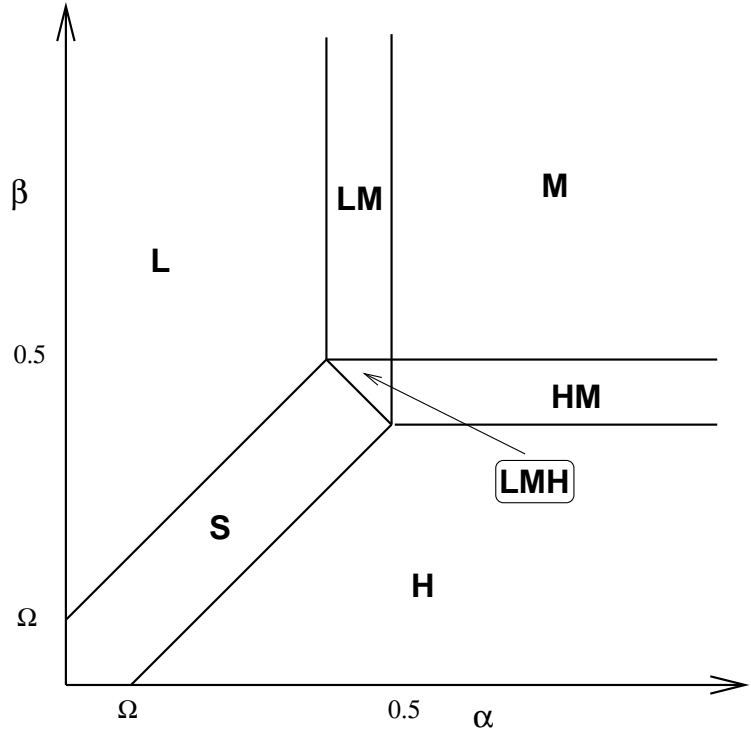,width=0.6\linewidth}}
\caption{Phase-diagram for $K=1$ and $\Omega < 1/2$. 
Indicated are 
the high- (H) and low-density (L) phase, the shock phase (S), the 
maximal current phase (M), and coexistence between a maximal current 
and low- (LM) or high-density (HM) domain. The LMH-phase indicates 
the presence of three domains. A linear density profile is observed at
the transition line between the LMH and S phase.}
\label{phase_k1}
\end{figure}

The above  results lead to the phase diagram shown in
fig.~\ref{phase_k1}. Three phases can be distinguished, where the 
density profile does not reach the equilibrium $\rho_{\rm eq}$: 
The high and low density phase, where only a single domain exists in the 
system, the shock (see fig.~\ref{fig_abI} for the corresponding
density profile) 
region, where high and low density domain coexist.  Then there are 
four phases, where the density profile has a section of  
constant density: The maximal current phase (M) is obtained for $\alpha > 1/2$
and $\beta > 1/2$. If $\alpha > 1/2$ and $1/2 > \beta > 1/2-\Omega$
we observe coexistence between a section of density $1/2$ at the left
and a linear profile as shown in fig.~\ref{fig_abIII}. This phase is indicated 
by (HM) in the phase diagram, the corresponding low density phase by 
(LM). Finally a section of constant density may coexist with 
a high- and low-density section (LMH) (see fig.~\ref{fig_abplateau}).
We observe all phases only if $\Omega < 1/2$. For  $1/2< \Omega < 1$
the high- and low density phases vanish, while for $\Omega > 1$
the formation of a shock is excluded.
Our MC analysis illustrates the validity of the mean field results for
large system. This is in contrast to the case $\Omega = 0$, where
e.g. for $\alpha = \beta$ the exact density profile even in the  
limit $N\to \infty$ is different from the mean field results.

\subsection{ The case  $K\neq1 $}
In the case $K\neq 1 $ the equilibrium  density is different from $\rho =
1/2$. Therefore it is impossible to observe a maximal current phase,
because the bulk particle reservoirs destroy any 
maximal-current domain. 
The absence of the maximal current phase originates in
(\ref{Char1},\ref{Char2}): For the kinematic wave to be stationary 
one requires both (\ref{Char1},\ref{Char2}) to be zero i.e. a bulk density
satisfying both
$\rho =  1/2$ and $\rho=\frac{K}{K+1}$. 
It is also important to notice that the 
solution of the mean field equations is not piecewise linear, i.e. higher 
order terms of eq.~(\ref{eq:dev}) do not vanish. We now analyse the mean field 
equations for $K>1$ in more detail. The corresponding results 
for $K< 1$ can be obtained from particle-hole symmetry. We consider
separately the regime where
$\alpha< 1/2$ and $\beta< 1/2$ and the complementary regime.\\

\noindent{\bf Case $\alpha< 1/2$ and $\beta< 1/2$ }

For low values of $\alpha$ and  $\beta $ we expect the existence of a
high-and low-density phase as well as the formation of shocks within a
certain density regime  as for the special case $K=1$. 
The transition lines can be obtained by analysing the shock
position, which is determined from the condition $\rho_r =
1-\rho_l$. The shock separates a region where the density 
is given by (\ref{lint}) and a region where the density is given by
 (\ref{rint}).

For low $\alpha$ the solution (\ref{lint}) may propagate 
all the way to the right boundary. If the density  at the right boundary 
satisfies $\rho_{l}(1) < \rho_{r}(1)=\beta$ any shock will be driven out
and the system will be dominated by $\rho_l$. 
This will be referred to as a low density region (LD).
The transition line between LD and S regions is given by
the condition
$\rho_{l}(1) =\beta$:
\be
\Omega_D(1+K)
&=&
\left[ 2(\beta-\alpha) + \frac{K-1}{(1+K)}
\ln \left| \frac{K-(1+K)\beta}{K-(1+K)\alpha}\right|\right] \ .
\label{LDSpb}
\ee

Similarly for large $\beta$ the solution $\rho_r$ may propagate all the way
to the left boundary where its density is
$\rho_r(0)$. A shock between $\rho_r$ and $\rho_l$
 will be driven to the left hand boundary if
$\rho_{r}(0) > 1-\rho_l(0)=1-\alpha$. This
will be referred to as the high density region (HD).
The transition line between HD and S regions is given by
the condition
$\rho_{r}(0) =1-\alpha$:
\be
\Omega_D(1+K)
&=&
\left[ 2(\alpha-\beta) + \frac{K-1}{(1+K)}
\ln \left| \frac{K-(1+K)(1-\beta)}{K-(1+K)(1-\alpha)}\right|\right] \ .
\label{HDSpb}
\ee
\vspace*{1ex}

\noindent{\bf Cases $\alpha >1/2$ or $\beta> 1/2$}

\begin{figure} 
\centerline{\epsfig{figure=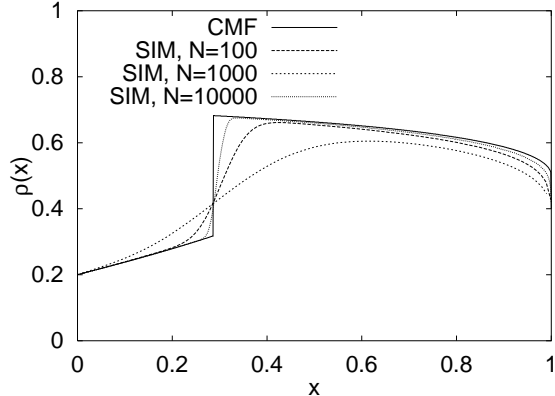,width=0.48\linewidth}}
\caption{
  Density profiles obtained from Monte Carlo simulations of the 
  model for $\alpha = 0.2$, $\beta = 0.6$, $K=3$, $\Omega_D = 0.1$
  and $N=100$, 1000, 10000.
  The full line gives  density profiles obtained from the continuum
  mean field theory, equations   (\ref{lint}) and (\ref{rint})
  with $\beta$ set equal to 1/2.}
\label{fig:k3dp}
\end{figure}

For $\alpha > 1/2$ or $\beta >1/2$ second-order terms
(e.g. terms involving $\partial^2 \rho/\partial x^2$) have to be
retained in (\ref{diffeq}) for steady state solutions to exist.
However the effect is, for example in the case $\alpha <1/2, \beta>
1/2$ at the right hand boundary, that over a finite distance (order
$1/N$ in terms of the variable $x$) the second order terms will match
the density with that implied by $\rho_r$. Thus the effect is that
there is no shock and $\beta >1/2$ may for the purposes of the phase
diagram may be {\em effectively} considered as $\beta
=1/2$. 
In fig.~\ref{fig:k3dp}
comparison of
direct simulations of the model when $\beta >1/2$
with density profiles obtained from the continuum
theory (with $\beta$
effectively considered as 1/2)
indicate the validity of our approach
for in the large $N$ limit.

For finite $N$,
finite size effects can partially be included by considering 
second-order terms in the mean-field description.
Deviations between second-order mean-field and 
the exact results are due to the fluctuations of the shock, which 
are not correctly described by mean-field theory and have to be treated 
separately. This will be done in the following section. 
However at this stage we note that the shock is indeed
localised and
that the width of the shock growth subextensively,
i.e. the shock is sharp in the limit $N\to \infty$.\\

\noindent{\bf Phase diagram}
\begin{figure} 
\centerline{\epsfig{figure=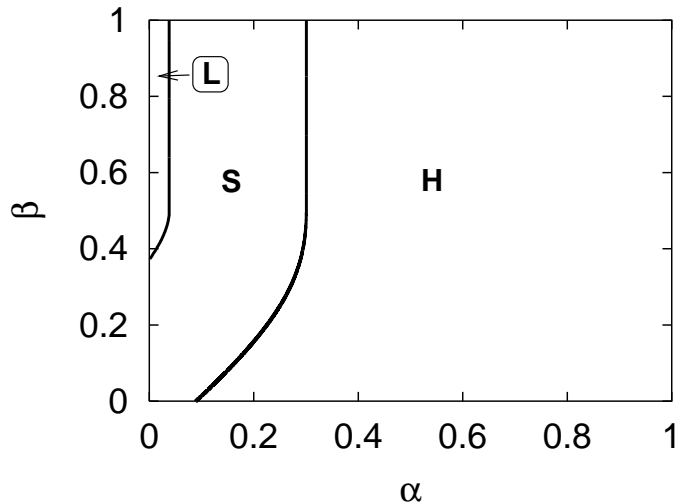,width=0.6\linewidth}}
\caption{Mean-field phase diagram for $\Omega_D = 0.1$  $K=3.$ (right). The phase 
boundaries between
the low density (L), shock (S) and high density (H) phases
 are calculated by using (\ref{LDSpb}) and (\ref{HDSpb}).
Our results are in excellent agreement with the findings of
 \cite{parma} who analysed numerically the second-order mean field 
equations.}
\label{fig:phasediagram}
\end{figure}

Our analysis for $K \neq 1$ leads to the phase diagram shown in
fig.~\ref{fig:phasediagram}.  We distinguish only three phases,
i.e. the high- and low-density phase as well as a the formation of
shocks in an intermediate parameter regime.  Compared to the case
$K=1$ the number of phases is considerably reduced, due to the absence
of a maximal current phase as discussed above.

As the shock is sharp
the fluctuations of the shock, which we treat in the next section,
do not affect the phase boundaries.
Therefore  we expect that the true phase diagram of the model  
is represented by fig.~\ref{fig:phasediagram}.

\section{Dynamics of the shock}
\label{shock}

A qualitative understanding of the shock dynamics can be easily obtained 
from mass conservation, i.e. by means of the continuity equation. 
Compared to the ASEP additional source and sink terms have to 
be introduced, reflecting the on-site input and output of
particles. This has been done in a recent work by Popkov et al. 
\cite{pop_prepint}, who generalised the domain wall picture to 
models with particle in- and output, where the continuity equation 
is given by:
\be 
\frac {\partial \rho(x,t)}{\partial t} +\frac{\partial }{\partial x}j(\rho)
= \Omega_A (1-\rho(x,t) )  - \Omega_D \rho(x,t)    \ .
\ee 
By inserting the known relation $j (\rho) =\rho \left ( 1-\rho \right )$
they recovered eq.~(\ref{diffeq}). The 
importance of the analysis by means of the hydrodynamic equation is
given by the fact, that this approach can be used for general models
if the flow-density relation is known \cite{pop_prepint}. 

\begin{figure}[h]
\centerline{\epsfig{figure=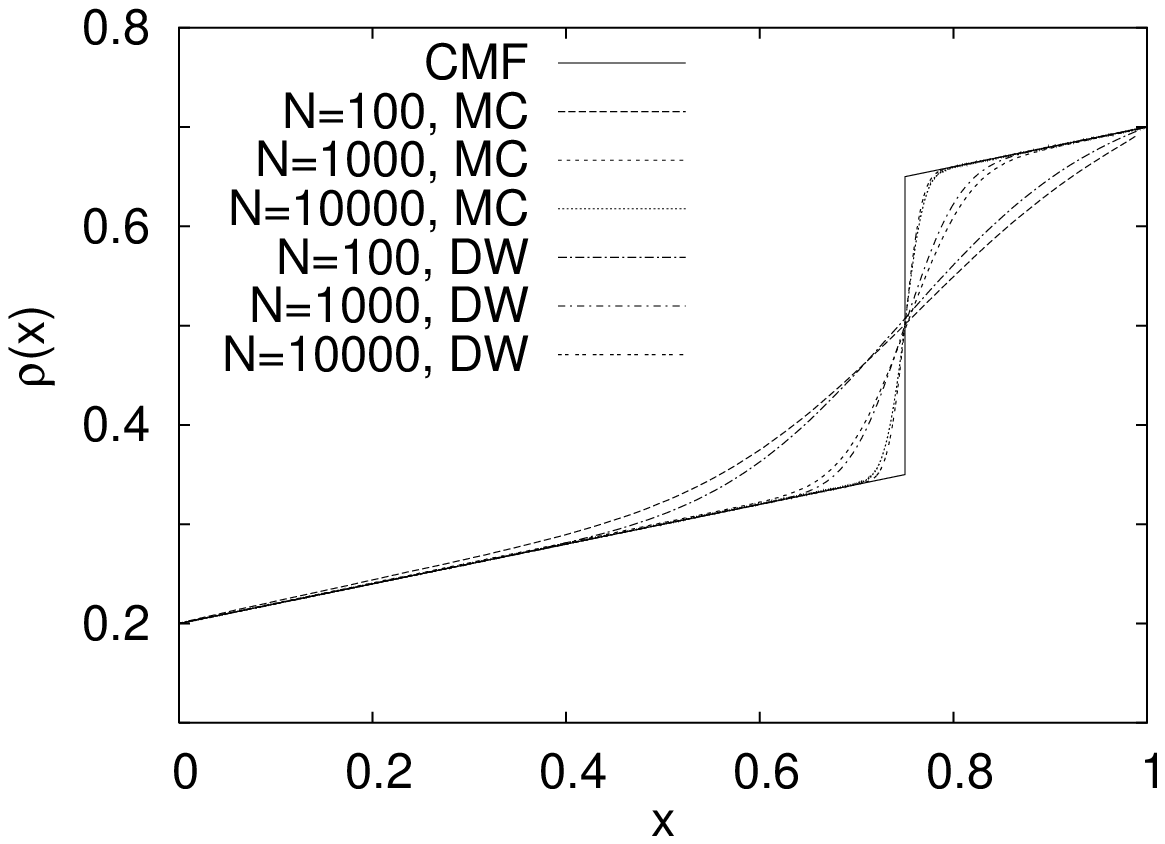,width=0.48\linewidth}\epsfig{figure=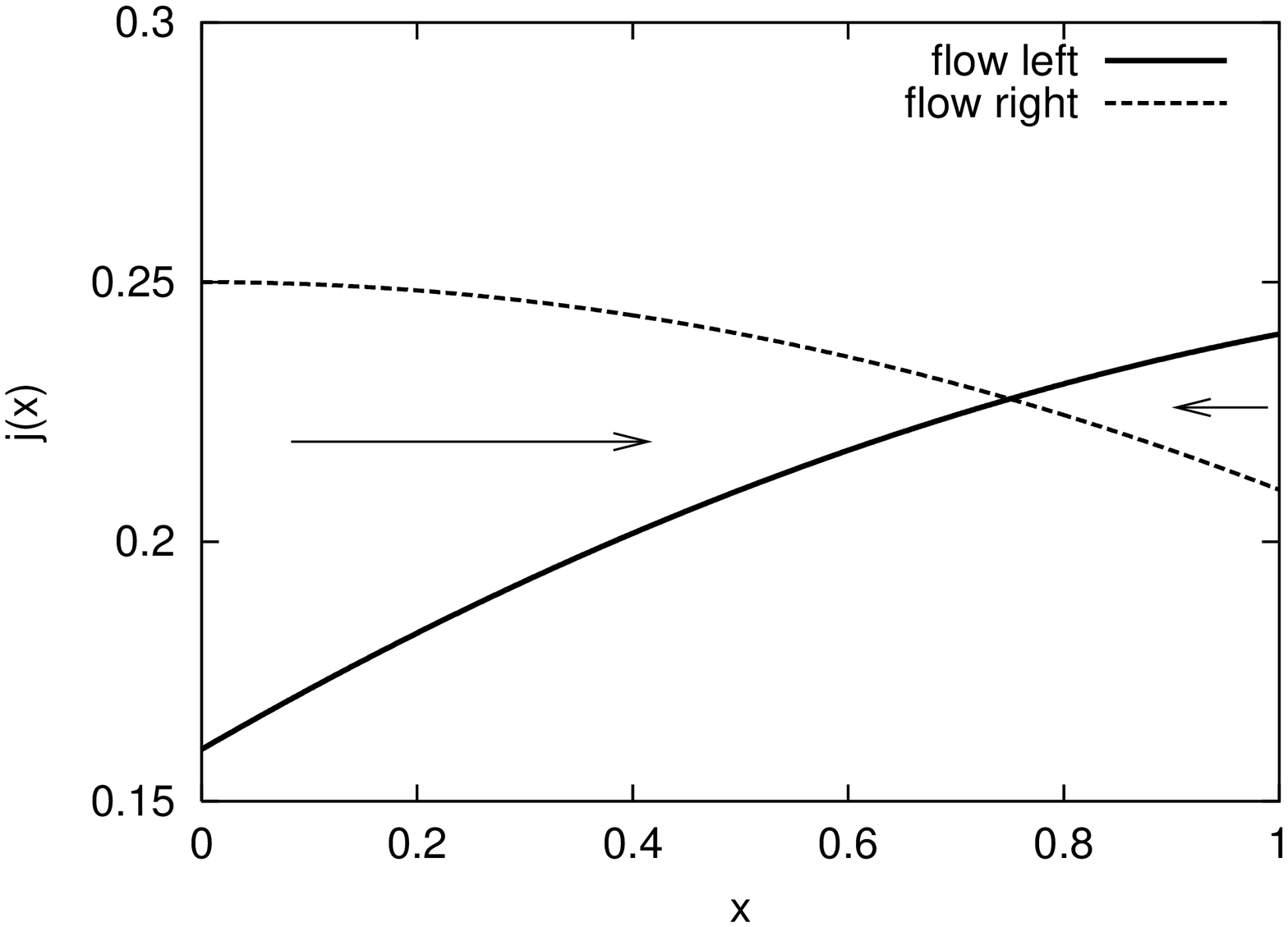,width=0.48\linewidth}}
\caption{Left: Density profiles in comparison with the
 DW predictions. The chosen parameters are $\alpha = 0.2$, 
 $\beta = 0.3$, $\Omega = 0.2$ and $K=1$. Right: Flow profile
 of the left (solid line) and right domain (dashed line) for same 
 set of parameters. The arrows indicate the bias of 
the random walker.}
\label{fig:localisation}
\end{figure}
A step beyond the mean field level is provided by interpreting the 
shock as a random walker.
For the ASEP on the line $\alpha=\beta$ the shock dynamics 
can be modelled as an  unbiased  random walk
\cite{DEM}
and this allows one to calculate its diffusion constant.
Moreover one can  generally consider kinematic
shock wave  dynamics as a biased random walker
\cite{ksks,sanapp}.
In the present case we 
model the shock by a random walker
with the site-dependent hopping rates
\be
w_l(i) = \frac{j_-(x)}{\Delta(i)} \qquad
w_r(i) = \frac{j_+(x)}{\Delta(i)},
\ee
where $w_l$ ($w_r$) denotes the hopping rate to the left (right), 
$j_-(i)$ $(j_+(i))$ the flux in the low (high) density domain at site $i$ 
and  $\Delta(i)$ the height of the shock at position $i$. Modelling 
the shock dynamics as a random walk with reflecting boundary conditions allows us to derive analytical 
expressions for the stationary distribution $p_s(i)$ of shock positions.
The stationary distribution $p_s(i)$ has to fulfil the condition 
\be 
 w_r (i) p_s(i) = w_l (i+1)p_s(i+1). 
\ee
One can solve this discrete equation explicitly, but it is simplest
to proceed by making a continuum approximation.
We expand the probability distribution to first order in 
$ 1/N$, as in eq.~(\ref{eq:dev}),
and use the stationarity condition to
obtain the differential equation:  
\be
y'(x)+ N y(x) \left ( 1- \frac{w_r(x)}{w_l(x)} \right )
= 0 \ ,  
\ee
 where $y(x) = p(x)w_l(x) $. The solution of this 
equation is given by 
\be
p(x) = \frac{\tilde{p}(x)}{\mathcal{N}} =
\frac{1}{\mathcal{N}w_l(x')}
\exp \left[- N \int_{x_0}^x
\left( 1-\frac{w_r(x')}{w_l(x')} \right ) dx'\right ] \ ,
\label{eq:shockdist}
\ee
where $\mathcal{N} = \int_0^1 \tilde{p}(x) dx$. Explicit expressions 
of the distribution can be given in case of $K=1$ which will be
discussed in detail. For this case we get the unnormalised distribution
\be
\tilde{p}(x) =
\left ( x+\frac{\alpha}{\Omega} \right )^{\frac{N(1-\Delta)\Delta}{\Omega}-1} 
\left ( \frac{1-\alpha}{\Omega} -x \right
)^{\frac{N(1+\Delta)\Delta}{\Omega}-1} \sim e^{-C (x-x_s)^2} 
 \ ,
\label{k1dist}
\ee
where $C = 4 N \Omega
\frac{\Delta}{(1-\Delta)(1+\Delta)}$. 
The Gaussian is obtained from a logarithmic 
expansion of the exponent, which is justified in the limit $N\to \infty$
\footnote{We also checked the accuracy of the Gaussian 
approximation by comparing with the numerically determined
exact distribution $p(x)$. The resulting density profiles 
are numerically indistinguishable for the system sizes shown in 
fig.~\ref{fig:localisation}.}. 

Using the random walk picture, the density profile can be obtained
from the shock distribution $p(x)$ through~\cite{sanapp} 
\be 
 \rho(x) = \rho_r(x) \int_0^x p(x') dx' + \rho_l(x) \int_x^1 p(x') dx'.
\ee 
The Gaussian approximation of $p(x)$ leads to 
 the following form of the density profile
\be 
 \rho(x) = \frac{\Delta}{2} \left [ 1 + \text{erf} \left ( 2 
   \sqrt{\frac{N \Omega \Delta}{(1+\Delta)(1-\Delta)}}(x-x_s) \right)
   \right ] + \Omega x + \alpha \ .
\label{eq_dp_dw}
\ee 
In order to check the validity of this simple phenomenological 
picture  we compare  the analytic predictions (\ref{eq_dp_dw}) for different 
density profiles in the (S) phase with results of MC
simulations \footnote{In order to pass the transient regime we
performed initially $10^5 L$ local updates. Measurements of the 
density profiles have been averaged over $10^8$  sweeps. The
statistical error of the MC data is of the order of the line 
width.}. Fig.~\ref{fig:localisation} shows that the DW predictions
are in good agreement  
with the MC results, and the accuracy of the DW description improves 
for larger system sizes.

From (\ref{eq_dp_dw}) it follows that the width of the shock scales 
as $N^{-\nu}$ with $\nu = 1/2$ as  was found numerically in \cite{parma}.

\section{Conclusion}
\label{conclude}
The model studied by Parmeggiani, Franosch and Frey \cite{parma} can be
interpreted as a generic model for the collective behaviour of
molecular motors. The physics of the model is governed by the
competition of different particle reservoirs and the stationary flow
of the self-driven particles. In the case of periodic boundary
conditions, one easily verifies that the stationary state of the
process is described by a product measure.  More interesting features
are observed in the case of open boundary conditions. Here, when the
rates for attachment and deletion of particles in the bulk are
appropriately scaled with system size, localisation of a shock arises
between the region of the system controlled by the left boundary and
the region controlled by the right boundary.

Most of the features of this model can be explained by a mean-field
analysis, which we believe to be correct in the limit of large
system size. This is supported by figs 1,2,3,5 where direct
simulations for the density profiles converge for 
large system sizes to our mean-field predictions.
In view of this  we believe
the mean-field phase diagrams figs 4,6 
are in fact exact.

By considering the characteristics of the
mean field equations the formation and localisation of the shock can
be explained: The characteristic solutions propagating from the left
and right boundaries are matched at the shock whose position is fixed
by the condition that the mass current through the shock is zero.  In
the presence of a shock, the leading finite-size corrections are due
to the fluctuations of the shock position, which can be described by
mapping the dynamics of the shock to a random walk with site dependent
hopping rates.

Apart from the importance of the ASEP with
creation and annihilation of particles in the bulk
as a  generic model for
molecular motors it is of special interest for the general formalism
of non-equilibrium statistical mechanics.  In particular
one can interpret the bulk non-conservation of particles as exchange of
particles with a bulk reservoir \cite{LKN}. In this way the  model can be
thought of as a grand-canonical counterpart to the ASEP.  As 
the ASEP and its variants can be analysed in detail, they might help us
to understand aspects of different ensembles in the context of
non-equilibrium statistical mechanics.

\begin{acknowledgments}
R.Juh{\'a}sz  and L.~Santen acknowledge  support from the Deutsche 
Forschungsgemeinschaft under Grant No. SA864/2-1.
MRE thanks Universit\"at des Saarlandes for kind hospitality  during a visit
supported  by  DFG (SFB277).
\end{acknowledgments}

\bibliography{Amine}

\begin{thebibliography}{25}

\bibitem{Liggett}
T. Liggett   \textit{Interacting particle systems: contact, voter
and exclusion processes} (Springer-Verlag: Berlin, 1999).

\bibitem{schutzreview}  G.M. Sch\"utz, in {\it Phase
  Transitions and Critical Phenomena}, vol. 19, Eds. C. Domb and J.L.
  Lebowitz (Academic Press, San Diego, 2001).

\bibitem{privman} V. Privman (ed.), 
{\it Nonequilibrium Statistical Mechanics in One Dimension}
 (Cambridge Univ. Press, Cambridge, 1997).
 
\bibitem{chowd} D. Chowdury, L.~Santen, A.~Schadschneider, 
 Phys. Rep. {\bf 329}, 199 (2000).

\bibitem{hinrichsen} H. Hinrichsen, 
Adv.~Phys.~{\bf 49}, 815 (2000).

\bibitem{schadschneider} A. Schadschneider,
Eur. Phys. J. B {\bf 10}, 573  (1999).

\bibitem{DEHP}
B. Derrida, M. R. Evans, V. Hakim,  and V. Pasquier 
J. Phys. A \textbf{26}, 1493 (1993).

\bibitem{SD}
G. Sch\"utz  and E. Domany,
J. Stat. Phys. {\bf 72}, 277 (1993).

\bibitem{ERS}
M. R. Evans, N. Rajewsky and E. R. Speer,
J. Stat. Phys. \textbf{95}, 45 (1999).

\bibitem{dGN}
J. de~Gier  and B. Nienhuis, 
Phys. Rev. E \textbf{59}, 4899 (1999).

\bibitem{rajewsky} N. Rajewsky, L. Santen, A. Schadschneider,
and  M. Schreckenberg, J. Stat. Phys. {\bf 92}, 151 (1998).

\bibitem{dls} B.~Derrida, J.L.~Lebowitz, and E.R.~Speer,
Phys. Rev. Lett. {\bf 89}, 030601 (2002). 

\bibitem{JL} For a  review see
S. A. Janowsky and J. L. Lebowitz  (1997)
chapter 13 in \cite{privman} and references therein.


\bibitem{parma}
A. Parmeggiani, T. Franosch, and E. Frey, 
\textit{Phase Coexistence in Driven One Dimensional
  Transport}, Phys.~Rev.~Lett {\bf 90}, 086601 (2003).

\bibitem{alberts_book} B. Alberts, D. Bray, J. Lewis, 
\textit{Molecular Biology of the Cell}, (Garland, New York, 1994).

\bibitem{howbook} J. Howard, 
\textit{Mechanics of Motor Proteins and the Cytoskeleton}, 
(Sinauer, Sunderland, 2001).

\bibitem{leshouches}
H.~Flyvbjerg, F.~J\"ulicher, P.~Ormos, F.~David (ed.),
\textit{Physics of Bio-Molecules and Cells}
(Springer, Heidelberg, 2002). 

\bibitem{howard} J. Howard, Nature  {\bf 389}, 561 (1997). 

\bibitem{juelrmp} F. J\"ulicher, A. Ajdari, and J. Prost,
Rev. Mod. Phys. {\bf 69}, 1269 (1997)

\bibitem{Davis} H. T. Davis 
\textit{Statistical Mechanics of Phases, Interfaces and Thin Films}
(Wiley-VCH, New York, 1996)

\bibitem{DDM}
B. Derrida, E.  Domany  and D. Mukamel 
J. Stat. Phys. \textbf{69}, 667  (1992).
 

\bibitem{Debnath} 
L.~Debnath,
\textit{Nonlinear Partial Differential Equations for Scientists and Engineers}
(Birkh\"auser, Boston, 1997)

\bibitem{LW} M.J. Lighthill and G.B. Whitham,
Proc. Roy. Soc. A {\bf 229}, 281 (1955). 

\bibitem{pop_prepint}  V. Popkov, A. R{\'a}kos, R. D. Willmann,
  A. B. Kolomeisky, and G. M. Sch\"utz, cond-mat/0302208
 (preprint)


\bibitem{DEM}
B. Derrida, M. R. Evans and K. Mallick,
J. Stat. Phys {\bf 79}, 833 (1995)


\bibitem{ksks} A. Kolomeisky, G.M. Sch\"utz, E.B. Kolomeisky, and
  J.P. Straley, J.~Phys. A {\bf 31}, 6911 (1998)


\bibitem{sanapp} 
 L.~Santen, and C. Appert, J.~Stat.~Phys. {\bf 106}, 187 (2002)  

\bibitem{LKN}
R. Lipowsky, S. Klumpp and T. M. Nieuwenhuizen,
Phys. Rev. Lett {\bf 87}, 108101 (2001); S. Klumpp
and R. Lipowsky cond-mat/0304681 (Preprint 2003)
\end{thebibliography}

\end{document}